\documentclass[a4paper,11pt]{article}
\usepackage{graphicx,amssymb,amsmath,amsfonts,epsfig}
\usepackage[usenames,dvipsnames]{color}
\usepackage[linktocpage,pagebackref]{hyperref}
\usepackage{subfigure}
\title{Energy Formula for Newman-Unti-Tamburino class of  Black Holes }
\author{Parthapratim Pradhan\footnote{pppradhan77@gmail.com}\\ 
\textit{Department of Physics}\\
\textit{Hiralal Mazumdar Memorial College For Women}\\
{Dakshineswar, Kolkata-700035, India}}
\date{\today}

\begin{document}

\maketitle

\begin{abstract}
We compute the \emph{surface energy~(${\cal E}_{s}^{\pm}$), 
the rotational energy~(${\cal E}_{r}^{\pm}$) and the electromagnetic energy~(${\cal E}_{em}^{\pm}$)} 
for Newman-Unti-Tamburino~(NUT) class of black hole having the event horizon~(${\cal H}^{+}$) 
and the Cauchy horizon~(${\cal H}^{-}$). Remarkably, we find that the 
\emph{mass parameter can be expressed as sum of  three energies i. e. 
$M={\cal E}_{s}^{\pm}+{\cal E}_{r}^{\pm}+{\cal E}_{em}^{\pm}$}. 
It has been \emph{tested} for Taub-NUT black hole, Reissner-Nordstr\"{o}m-Taub-NUT 
black hole, Kerr-Taub-NUT black hole and Kerr-Newman-Taub-NUT black hole. 
In each case of black hole, we find that \emph{the sum of these energies is equal to the Komar mass}. 
It is plausible only due to the introduction of new conserved charges i.e. 
$J_{N}=M\,N$~(where $M=m$ is the Komar mass and $N=n$ is the gravitomagnetic charge), 
which is closely analogue to the Kerr-like angular momentum parameter $J=a\,M$.
\end{abstract}

\newpage

\textheight 25 cm

\clearpage

\section{Introduction}
In recent times there is considerable ongoing interest in the thermodynamics of 
the NUT class of black holes. Much of the insight gained 
from the work~\cite{wu} in which it has been suggested that a generic four 
dimensional Taub-NUT black hole~(BH) should be interpreted as in terms of 
three or four different types of thermodynamic hairs.  They could be 
defined as the Komar mass~($M=m$), the angular momentum~($J_{n}=mn$), 
the gravitomagnetic charge~($N=n$),  the dual~(magnetic) mass 
$(\tilde{M}=n)$. 

Motivated by this formalism, in the present work, we wish to  compute \emph{the 
surface energy~(${\cal E}_{s}^{\pm}$), the rotational energy~(${\cal E}_{r}^{\pm}$) and the 
electromagnetic energy~(${\cal E}_{em}^{\pm}$))} for NUT class of BH having two 
physical horizons i. e., the event horizon~(${\cal H}^{+}$) and 
the Cauchy horizon~(${\cal H}^{-}$). Long ago, Smarr~\cite{smarr73a,smarr73b} computed these 
energies for three-parameter, charged Kerr BH, which is a solution of the Einstein-Maxwell equation. 
The author calculated the area of  the Kerr-Newman~(KN) BH, and from this expression he
 derived the mass parameter as a function of the area, angular momentum, and charge. 
He also proved the mass differential could be expressed as three physical invariants of the BH horizon. 
These three physical invariants are effective surface 
tension~(${\cal T}$), angular velocity~($\Omega$), and electromagnetic potential~($\Phi$). 
Moreover, he demonstrated that the mass parameter should be expressed in terms of these physical
invariants as a simple bilinear form. 
This expression could be derived by applying Euler's theorem on homogeneous functions to the 
mass parameter. 

Furthermore, he showed that the effective surface tension, the angular velocity, and 
the electromagnetic potentials may be defined and are constant on the horizon for any 
stationary axisymmetric spacetime. 
Also, he evaluated three energy components for charged Kerr BH, i. e. 
the surface energy, the rotational energy, and electromagnetic energy. 
Further, these integrals are computed using the variational definitions. 
Then it was easily shown that the sum of three energies is equal to the
mass parameter. 

There are several important works we should mention here that has been discussed in 
different aspects of NUT BH~\cite{taub51,mis63,mis63a,mis67,mis69,mkg71,miller,sen81,bell97,chur,chur1,hunter98,hunter99,page99,myers99,empa99}. 
If we have not considered the formalism developed recently in~\cite{wu}, the following  
relations do exist~\cite{pp15,pp16,pp16a}: \\
(i) the  effective surface tension of  ${\cal H}^{\pm}$ is \emph{not} proportional 
to the surface gravity of ${\cal H}^{\pm}$  i. e.
$$
{\cal T}_{\pm} \neq \frac{{\kappa}_{\pm}} {8\pi},
$$
where $\kappa_{\pm}$ is the surface gravity of ${\cal H}^{\pm}$.\\
(ii) the mass parameter can not be expressed in terms of three physical invariants, 
i.e., the effective surface tension of ${\cal H}^{\pm}$, the angular velocity of ${\cal H}^{\pm}$ 
and the electromagnetic potential of ${\cal H}^{\pm}$\\
$$
M \neq 2{\cal T}_{\pm}\, {\cal A}_{\pm} + 2J_{N}\, \omega_{\pm} +\psi_{\pm}\, N, 
$$
and finally, \\
(iii) the sum of the surface energy, the rotational energy, and the electromagnetic energy is not 
equal to the mass parameter i.e.
$$
{\cal E}_{s}^{\pm}+{\cal E}_{r,\,J_{N}}^{\pm}+{\cal E}_{em,\,J_{N}}^{\pm} \neq M
$$
In the \emph{present work}, we want to calculate \emph{three energy components, 
i.e., the surface energy of ${\cal H}^{\pm}$, the rotational energy of ${\cal H}^{\pm}$ 
and electromagnetic energy of ${\cal H}^{\pm}$ } for NUT class of BH. Notably, we point 
out that the \emph{sum of these energies is equal to the mass parameter}.

Introducing the formalism developed in~\cite{wu} we are 
indeed able to \emph{prove that}\\
(i) the  effective surface tension of  ${\cal H}^{\pm}$ is proportional 
to the surface gravity of ${\mathcal H}^{\pm}$  i. e.
$$
{\cal T}_{\pm} = \frac{{\kappa}_{\pm}} {8\pi},
$$
\\
(ii) the mass parameter can  be expressed in terms of three physical invariants i.e. 
$$
 M = 2{\cal T}_{\pm}\, {\cal A}_{\pm} + 2J_{N}\, \omega_{\pm} +\psi_{\pm}\, N, 
$$
and finally, 
\\
(iii) the \emph{sum of three energies is equal to the mass parameter} i.e.  
$$
{\cal E}_{s}^{\pm}+{\cal E}_{r,\,J_{N}}^{\pm}+{\cal E}_{em,\,J_{N}}^{\pm} = M
$$
We have explicitly examined the result~(iii) especially for the NUT class of 
BHs, i.e., for Taub-NUT~(TN) BH, Reissner-Nordstr\"{o}m Taub-NUT BH, Kerr-Taub-NUT BH 
and Kerr-Newman-Taub-NUT BH. We show that \emph{the sum of three energies is indeed equal 
to the mass parameter}. It has not been studied previously to the best of my knowledge. 
It is possible only due to the introduction of new conserved charges, i.e., $J_{N}=M\,N$ which 
is analogous to the Kerr like  angular 
momentum parameter $J=a\,M$. The other reason is that due to multihair 
features of  NUT parameter~i.e. both rotation-like and electromagnetic 
charge like characteristics:
$$\frac{J_{n}}{m}=n\equiv N=\frac{J_{N}}{M}$$.

In the next Sec.~(\ref{1a}), we have derived the surface energy, the rotational energy, and 
the electromagnetic energy for Taub-NUT BH. In Sec.~(\ref{2b}), we have derived these energies
for Reissner-Nordstr\"{o}m-Taub-NUT BH and proved that the sum of three energies equal to the 
mass parameter. A similar analysis has been done for Kerr-Taub-NUT BH in 
Sec.~(\ref{3c}). In Sec.~(\ref{3d}), we have done a similar analysis for Kerr-Newman-Taub-NUT BH. 
Finally, we have given our conclusions in  Sec.~(\ref{con}).

\section{\label{1a} Energy Formula for Taub-NUT BH}
First, we consider the Taub-NUT BH. The metric~\cite{mis63,mis63a,mis67,mis69,mkg71} in 
Schwarzschild like coordinates are  
\begin{eqnarray}
ds^2 &=& -{\cal Y}(r) \, \left(dt+2n\cos\theta d\phi\right)^2+ \frac{dr^2}{{\cal Y}(r)}
+\left(r^2+n^2\right) \left(d\theta^2+\sin^2\theta d\phi^2 \right) ~,\label{tn}
\end{eqnarray}
where the function ${\cal Y}(r)$ is defined by 
\begin{eqnarray}
 {\cal Y}(r) &=& \frac{1}{r^2+n^2} \left[r^2-n^2-2mr \right]
\end{eqnarray}
Under new procedure~\cite{wu} for NUT class of BHs, the global 
conserved charges are defined as 
\begin{eqnarray}
\mbox{Komar mass} :\, M &=& m, ~\nonumber\\
\mbox{Gravitomagnetic charge}: \, N &=& n, ~\nonumber\\
\mbox{Angular momentum}: \, J_{n} &=& m\,n, ~\label{pn}\\
\mbox{Dual~(or Magnetic) mass}: \, \tilde{M} &=& n \equiv N. ~\label{td}
\end{eqnarray}
Taking  cognizance of Eq.~({\ref{pn}}) then the metric can be re-written as 
\begin{eqnarray}
ds^2 &=& -{\cal Y}(r) \, \left(dt+2N\,\cos\theta d\phi\right)^2+ \frac{dr^2}{{\cal Y}(r)}
+\left(r^2+N^2\right)\left(d\theta^2+\sin^2\theta d\phi^2 \right) ~,\label{tn1}
\end{eqnarray}
where the function ${\cal Y}(r)$ is given by 
\begin{eqnarray}
{\cal Y}(r) &=& \frac{1}{r^2+N^2} \left(r^2-N^2-2Mr\right)
\end{eqnarray}
The Killing horizons are located at 
\begin{eqnarray}
r_{\pm}= M \pm \sqrt{M^2+N^2}.
\end{eqnarray}
$r_{+}$ is called EH  and $r_{-}$  is called CH. Now we know from Hawking's~\cite{bcw73}
general theorem which asserted  that the surface area of a BH never decrease. Thus for a 
Taub-NUT class of BH, the area is constant. It is derived to be both for the 
horizons~\cite{pp20} as
\begin{eqnarray}
{\cal A}_{\pm}  &=&  8\pi\left[M^2+N^2 \pm \sqrt{M^4+J_{N}^2}\right]
~.\label{tn2}
\end{eqnarray}
According to Ramaswamy and Sen~\cite{sen86}, it was pointed out 
that NUT solutions with the mass parameter $M=0$ i.e. massless dual mass are perfectly well defined. 
In this case, the horizon radius becomes $r_{\pm}= \pm N$. Consequently, the area~[Eq.~(\ref{tn2})] 
of both the horizons reduced to ${\cal A}_{\pm}=8\pi N^2$. Hence in this circumtances,  
only the NUT parameter can be expressed as in terms of the area of both the horizons 
i.e.
\begin{eqnarray}
N=\sqrt{\frac{{\cal A}_{\pm}}{8\pi}}. 
\end{eqnarray}
In this situation the mass formula does not exist.

On inverting Eq.~(\ref{tn2}), one obtains the Komar mass as 
a function of area, new conserved charges $J_{N}$ and NUT 
parameter for both the horizons ${\cal H}^{\pm}$:
\begin{eqnarray}
M\,({\cal A}_{\pm}, J_{N}, N) &=& 
\sqrt{\frac{{\cal A}_{\pm}}{16\pi}+\frac{4\pi\,J_{N}^2}{{\cal A}_{\pm}}-N^2+\frac{4\pi\,N^4}{{\cal A}_{\pm}}},
~\label{tn3}
\end{eqnarray}
It is interesting to note that the mass parameter can be expressed in terms of both the 
area of ${\cal H}^+$ and ${\cal H}^-$. The mass differential is expressed as three
physical invariants of both ${\cal H}^+$ and ${\cal H}^-$
\begin{eqnarray}
dM &=& {\cal T}_{\pm}\, d{\cal A}_{\pm} + \omega_{\pm}\, dJ_{N} +\psi_{\pm}\,dN
~. \label{tn4}
\end{eqnarray}
where
\begin{eqnarray}
{\cal T}_{\pm} &=&  \frac{1}{{ M}} \left(\frac{1}{32 \pi}-\frac{2\pi J_{N}^2}{{\cal A}_{\pm}^2}-\frac{2\pi N^4}
{{\cal A}_{\pm}^2} \right)~.\label{tn5} \\
\omega_{\pm} &=& \frac{4\pi J_{N}}{M{\cal A}_{\pm}}, ~.\label{tn6} \\
\psi_{\pm} &=& -\frac{N}{{M}} \left(1-\frac{8\pi N^2}{{\cal A}_{\pm}}\right)  ~. \label{tn7}
\end{eqnarray}
and
\begin{eqnarray}
{\cal T}_{\pm} &=& \mbox{Effective surface tension of ${\cal H}^+$ and ${\cal H}^-$} \nonumber \\
\omega_{\pm} &=&  \mbox{Angular velocity of ${\cal H}^\pm$} \nonumber \\
\psi_{\pm} &=& \mbox{NUT potentials of ${\cal H}^\pm$}\nonumber
\end{eqnarray}

Shortly, we prove that this effective surface tension is proportional to the surface gravity 
of the horizons, i.e.
\begin{eqnarray}
{\cal T}_{\pm} &=& \frac{1}{M} \left[\frac{1}{32 \pi}-\frac{2\pi J_{N}^2}
                   {{\cal A}_{\pm}^2}-\frac{2\pi N^4}{{\cal A}_{\pm}^2} \right] \\
&=& \frac{1}{32 \pi { M}} \left[1-\frac{64\pi^2(J_{N}^2+N^4)}{{\cal A}_{\pm}^2}\right]\nonumber\\
&=& \frac{1}{16 \pi { M}} \left[1-\frac{({ M}^2+N^2)}{M r_{\pm}+N^2}\right]\nonumber\\
&=& \pm \frac{\sqrt{{ M}^2+N^2}}{8\pi\,\left(r_{\pm}^2+N^2 \right)} \nonumber\\
&=& \frac{r_{\pm}-{M}}{8\pi\,\left(r_{\pm}^2+N^2 \right)}= \frac{{\kappa}_{\pm}} {8\pi},~\label{tn8}
\end{eqnarray}
where $\kappa_{\pm}$ is defined {as} the surface gravity~\cite{bk73,bcw73} of ${\cal H}^{\pm}$. 
The BH temperature of both the horizons {is} derived via surface gravity on the horizons  
\begin{equation}
T_{\pm}=\frac{\kappa_{\pm}}{ 2\pi} 
=\frac{r_{\pm}-M}{2\pi\left(r_{\pm}^2+N^2\right)}
=\frac{1}{4\pi\,r_{\pm}}. ~\label{tn9}
\end{equation}
The entropy can be easily derived from the area formula via this relation.
$$
S_{\pm}=\frac{A_{\pm}}{4}
$$
So, the mass parameter~\cite{wu} should be expressed in terms of these quantities both 
for ${\cal H}^\pm$ as a simple bilinear form
\begin{eqnarray}
 M &=& 2{\cal T}_{\pm}\, {\cal A}_{\pm} + 2J_{N}\, \omega_{\pm} +\psi_{\pm}\, N
~. \label{tn10}
\end{eqnarray}
This striking formula is obtained by applying Euler's theorem on homogeneous functions 
to $M$, which is homogeneous of degree $\frac{1}{2}$ in $({\cal A}_{\pm},\,J_{N},\, N^2)$.
This may be rewritten as 
\begin{eqnarray}
 M &=& \frac{{\kappa}_{\pm}} {4\pi}\, {\cal A}_{\pm} + 2J_{N}\, \omega_{\pm} +\psi_{\pm}\, N ~. \label{tn11}
\end{eqnarray}
Therefore  the \emph{Smarr-Gibbs-Duhem} relation of ${\cal H}^{\pm}$ for Taub-NUT BH is 
\begin{eqnarray}
\frac{{M}}{2} &=& {T}_{\pm}{\cal S}_{\pm} + J_{N}\,\omega_{\pm}+\frac{\psi_{\pm}\,N}{2}
~. \label{tn14}
\end{eqnarray}
Interestingly, ${\cal T}_{\pm}$,  $\omega_{\pm}$ and $\psi_{\pm}$ should be defined 
and are constant on the ${\cal H}^+$ and ${\cal H}^-$ for any stationary, axially 
symmetric space-time.
Since  $dM$ is a perfect differential therefore one can choose freely any path of
integration in $({\cal A}_{\pm},\, J_{N},\, N)$ space. So the surface energy 
${\cal E}_{s}^{\pm}$ of ${\cal H}^+$~\cite{smarr73a} and ${\cal H}^{-}$~\cite{pp14} 
may be defined by
\begin{eqnarray}
{\cal E}_{s}^{\pm} &=& \int_{0}^{{\cal A}_{\pm}} {\cal T}_{\pm} \,(\tilde{{\cal A}_{\pm}}
, 0 ,0)\, d\tilde{{\cal A}_{\pm}}; ~ \label{tn15}
\end{eqnarray}
the rotational energy  of ${\cal H}^+$~\cite{smarr73a} and ${\cal H}^{-}$~\cite{pp14} can be defined by
\begin{eqnarray}
{\cal E}_{r,\,J_{N}}^{\pm} &=& \int_{0}^{J_{N}} \omega_{\pm}\, ({\cal A}_{\pm}
, \tilde{J}_{N} ,0)\, d\tilde{J}_{N},\,\,  \mbox{${\cal A}_{\pm}$ fixed}; ~ \label{tn16}
\end{eqnarray}
and the electromagnetic energy  of ${\cal H}^+$~\cite{smarr73a} and ${\cal H}^{-}$~\cite{pp14} can be 
defined as 
\begin{eqnarray}
{\cal E}_{em,\,J_{N}}^{\pm} &=& \int_{0}^{N} \psi_{\pm} \,({\cal A}_{\pm}
, J_{N}, \tilde{N})\,d\tilde{N},\,\, \mbox{${\cal A}_{\pm}$, $J_{N}$  fixed};  ~ \label{tn17}
\end{eqnarray}
These integrals should be directly computed using the variational definitions which is already 
defined in Eqs.~({\ref{tn5}},{\ref{tn6}},{\ref{tn7}}).
First we shall compute the surface energy of ${\cal H}^{\pm}$:
\begin{eqnarray}
{\cal E}_{s}^{\pm} &=& \int_{0}^{{\cal A}_{\pm}} {\cal T}_{\pm} \,(\tilde{{\cal A}_{\pm}}
, 0 ,0)\, d\tilde{{\cal A}_{\pm}}\\
                   &=& \sqrt{\frac{{\cal A}_{\pm}}{16\pi}} ~ \label{tn18}
\end{eqnarray}
Next we find the rotational energy of ${\cal H}^{\pm}$ as
\begin{eqnarray}
{\cal E}_{r,\,J_{N}}^{\pm} &=& \int_{0}^{J_{N}} \omega_{\pm}\, ({\cal A}_{\pm}
, \tilde{J}_{N} ,0)\, d\tilde{J}_{N},\,\,  \mbox{${\cal A}_{\pm}$ fixed}\\
                   &=& \sqrt{\frac{{\cal A}_{\pm}}{16\pi}+\frac{4\pi\,J_{N}^2}{{\cal A}_{\pm}}}
                       -\sqrt{\frac{{\cal A}_{\pm}}{16\pi}}.~ \label{tn19}
\end{eqnarray}
and finally we get the electromagnetic energy  of ${\cal H}^{\pm}$  as 
\begin{eqnarray}
{\cal E}_{em,\,J_{N}}^{\pm} &=& \int_{0}^{N} \psi_{\pm} \,({\cal A}_{\pm}
, J_{N}, \tilde{N})\,d\tilde{N},\,\, \mbox{${\cal A}_{\pm}$, $J_{N}$ fixed}\\
     &=& \sqrt{\frac{{\cal A}_{\pm}}{16\pi}+\frac{4\pi\,J_{N}^2}{{\cal A}_{\pm}}-N^2+\frac{4\pi\,N^4}{{\cal A}_{\pm}}} 
     -\sqrt{\frac{{\cal A}_{\pm}}{16\pi}+\frac{4\pi\,J_{N}^2}{{\cal A}_{\pm}}}.~ \label{tn20}
\end{eqnarray}
Now we compute the sum of three energies 
\begin{eqnarray}
{\cal E}_{s}^{\pm}+{\cal E}_{r,\,J_{N}}^{\pm}+{\cal E}_{em,\, J_{N}}^{\pm} &=&  
\sqrt{\frac{{\cal A}_{\pm}}{16\pi}+\frac{4\pi\,J_{N}^2}{{\cal A}_{\pm}}-N^2+\frac{4\pi\,N^4}{{\cal A}_{\pm}}}.
~ \label{tn21}
\end{eqnarray}
Using Eq.~({\ref{tn3}}), we can rewrite the above equation as
\begin{eqnarray}
{\cal E}_{s}^{\pm}+{\cal E}_{r,\,J_{N}}^{\pm}+{\cal E}_{em,\,J_{N}}^{\pm} &=&  
\sqrt{\frac{{\cal A}_{\pm}}{16\pi}+\frac{4\pi\,J_{N}^2}{{\cal A}_{\pm}}-N^2+\frac{4\pi\,N^4}{{\cal A}_{\pm}}}=
M\,({\cal A}_{\pm}, J_{N}, N) .~ \label{tn22}
\end{eqnarray}
Remarkably, the Komar mass can be expressed as the sum of three energies, namely the surface energy of 
${\cal H}^{\pm}$, the rotational energy of ${\cal H}^{\pm}$ 
and the electromagnetic energy of ${\cal H}^{\pm}$. 

Note that, in the limit $N=0$, we get the energy formula for Schwarzschild BH i.e.
\begin{eqnarray}
{\cal E}_{s}^{\pm} &=&  \sqrt{\frac{{\cal A}_{\pm}}{16\pi}}= M\,({\cal A}_{\pm}) .~ \label{tn22.1}
\end{eqnarray}
In the next section, we will do a similar investigation by adding a charge parameter.

\section{\label{2b} Energy formula for Reissner-Nordstr\"{o}m-Taub-NUT BH}
Now we want to extend the preceeding analysis for Reissner-Nordstr\"{o}m-Taub-NUT BH. The metric 
function modified for this BH as 
\begin{eqnarray}
 {\cal Y}(r) &=& \frac{1}{r^2+N^2} \left(r^2-N^2-2Mr+Q^2 \right)
\end{eqnarray}
where $Q$ is the purely electric charge. Now the BH horizons are situated  at 
\begin{eqnarray}
r_{\pm}= M \pm\sqrt{ M^2-Q^2+N^2}
\end{eqnarray}
Now the area of the BH computed for both the physical horizons are 
\begin{eqnarray}
{\cal A}_{\pm} &=&  4\pi\left[2 \left(M^2+N^2 \right)-Q^2 \pm 2\sqrt{M^4-M^2Q^2+J_{N}^2}\right] ~.\label{b1}
\end{eqnarray}
Now inverting the above expression, one obtains the Komar mass as 
a function of area,  conserved charges $J_{N}$, charge parameter $Q$ and 
NUT parameter~($N$) for  ${\cal H}^{\pm}$:
\begin{eqnarray}
M\,({\cal A}_{\pm}, J_{N}, N, Q) &=& 
\sqrt{\frac{{\cal A}_{\pm}}{16\pi}+\frac{4\pi\,J_{N}^2}{{\cal A}_{\pm}}-N^2+\frac{4\pi\,N^4}{{\cal A}_{\pm}}
+\frac{\pi Q^4}{{\cal A}_{\pm}}-\frac{4\pi \,N^2\,Q^2}{{\cal A}_{\pm}}+\frac{Q^2}{2}},\nonumber \\
~\label{b3}
\end{eqnarray}
So,  the mass differential can be expressed as three
physical invariants of both ${\cal H}^{\pm}$
\begin{eqnarray}
dM &=& {\cal T}_{\pm}\, d{\cal A}_{\pm} + \omega_{\pm}\, dJ_{N} +\psi_{\pm}\,dN+\Phi_{\pm}\,dQ
~. \label{b4}
\end{eqnarray}
where
\begin{eqnarray}
{\cal T}_{\pm} &=&  \frac{1}{{ M}} \left(\frac{1}{32 \pi}-\frac{2\pi J_{N}^2}{{\cal A}_{\pm}^2}-\frac{2\pi N^4}
{{\cal A}_{\pm}^2}-\frac{\pi\,Q^4}{2{\cal A}_{\pm}^2}+\frac{2\pi \,N^2\,Q^2}{{\cal A}_{\pm}^2} \right)~.\label{b5} \\
\omega_{\pm} &=& \frac{4\pi J_{N}}{M{\cal A}_{\pm}}, ~.\label{b6} \\
\psi_{\pm} &=& -\frac{N}{{M}} \left(1-\frac{8\pi N^2}{{\cal A}_{\pm}}+\frac{4\pi\,Q^2}{{\cal A}_{\pm}}\right)  ~. \label{b7}\\
\Phi_{\pm} &=& \frac{1}{M} \left(\frac{2\pi\,Q^3}{{\cal A}_{\pm}}-\frac{4\pi\,Q\,N^2}{{\cal A}_{\pm}}+\frac{Q}{2} \right)
 ~. \label{b7.1}
\end{eqnarray}
and
\begin{eqnarray}
{\cal T}_{\pm} &=& \mbox{Effective surface tension of ${\cal H}^+$ and ${\cal H}^-$ for RN-Taub-NUT BH} \nonumber \\
\omega_{\pm} &=&  \mbox{Angular velocity of ${\cal H}^\pm$ for RN-Taub-NUT BH} \nonumber \\
\psi_{\pm} &=& \mbox{NUT potentials of ${\cal H}^\pm$ for RN-Taub-NUT BH}\nonumber\\
\Phi_{\pm} &=& \mbox{Electromagnetic potentials of ${\cal H}^\pm$ for RN-Taub-NUT BH}\nonumber
\end{eqnarray}

Similarly, we show  that this effective surface tension is proportional to the surface gravity 
of the horizons, i.e.
\begin{eqnarray}
{\cal T}_{\pm} &=& \frac{r_{\pm}-{M}}{8\pi\,\left(r_{\pm}^2+N^2 \right)}
            =\frac{{\kappa}_{\pm}} {8\pi},~\label{b8}
\end{eqnarray}
where $\kappa_{\pm}$ is defined as the surface gravity of ${\cal H}^{\pm}$ for RN-Taub-NUT BH. 

Hence, for RN-Taub-NUT BH, the mass parameter can be expressed in terms of these quantities both 
for ${\cal H}^\pm$ as a simple bilinear form
\begin{eqnarray}
 M &=& 2{\cal T}_{\pm}\, {\cal A}_{\pm} + 2J_{N}\, \omega_{\pm} +\psi_{\pm}\, N+\Phi_{\pm}\, Q
~. \label{b10}
\end{eqnarray}
Like Taub-NUT BH, this striking formula could be obtained by applying Euler's 
theorem on homogeneous functions to $M$, which is homogeneous of degree 
$\frac{1}{2}$ in $({\cal A}_{\pm},\,J_{N},\, N^2, Q^2)$.
Thus this can be rewritten as 
\begin{eqnarray}
M &=& \frac{{\kappa}_{\pm}} {4\pi}\, {\cal A}_{\pm} + 2J_{N}\, \omega_{\pm} +\psi_{\pm}\, N+\Phi_{\pm}\,Q
 ~. \label{b11}
\end{eqnarray}

Thus  the \emph{Smarr-Gibbs-Duhem} relation of ${\cal H}^{\pm}$ for RN-Taub-NUT BH is 
\begin{eqnarray}
\frac{{M}}{2} &=& {T}_{\pm}{\cal S}_{\pm} + J_{N}\,\omega_{\pm}+\frac{\psi_{\pm}\,N}{2}+\frac{\Phi_{\pm}\,Q}{2}
~. \label{b14}
\end{eqnarray}
 Interestingly, ${\cal T}_{\pm}$,  $\omega_{\pm}$, $\psi_{\pm}$ and $\Phi_{\pm}$ could be  defined 
and are constant on the ${\cal H}^{\pm}$ for RN-Taub-NUT spacetime.

As  $dM$ is a perfect differential therefore one can choose freely any path of
integration in $({\cal A}_{\pm},\, J_{N},\, N, Q)$ space. So the surface energy 
${\cal E}_{s}^{\pm}$ of ${\cal H}^{\pm}$ for RN-Taub-NUT BH can be defined by
\begin{eqnarray}
{\cal E}_{s}^{\pm} &=& \int_{0}^{{\cal A}_{\pm}} {\cal T}_{\pm} \,(\tilde{{\cal A}_{\pm}}
, 0 ,0,0)\, d\tilde{{\cal A}_{\pm}}; ~ \label{b15}
\end{eqnarray}
The rotational energy  of ${\cal H}^{\pm}$ can be defined by
\begin{eqnarray}
{\cal E}_{r,\,J_{N}}^{\pm} &=& \int_{0}^{J_{N}} \omega_{\pm}\, ({\cal A}_{\pm}
, \tilde{J}_{N} ,0,0)\, d\tilde{J}_{N},\,\,  \mbox{${\cal A}_{\pm}$ fixed}; ~ \label{b16}
\end{eqnarray}
The electromagnetic energy  of ${\cal H}^{\pm}$ due to NUT parameter~($N$) can be defined as 
\begin{eqnarray}
{\cal E}_{em,\,J_{N}}^{\pm} &=& \int_{0}^{N} \psi_{\pm} \,({\cal A}_{\pm}
, J_{N}, \tilde{N},0)\,d\tilde{N},\,\, \mbox{${\cal A}_{\pm}$, $J_{N}$  fixed};  ~ \label{b17}
\end{eqnarray}
and the electromagnetic energy  of ${\cal H}^{\pm}$ due to  charge parameter~($Q$) 
can be defined as 
\begin{eqnarray}
{\cal E}_{em,\,J_{N},\,Q}^{\pm} &=& \int_{0}^{Q} \Phi_{\pm} \,({\cal A}_{\pm}
, J_{N}, N, \tilde{Q})\,d\tilde{Q},\,\, \mbox{${\cal A}_{\pm}$, $J_{N}$, $N$ fixed};~ \label{b18}
\end{eqnarray}
These integrals should be directly computed using the variational definitions which is already 
defined in Eqs.~({\ref{b5}}),({\ref{b6}}),({\ref{b7}}),({\ref{b7.1}}).

Proceeding analogously the surface energy of ${\cal H}^{\pm}$ for RN-Taub-NUT BH is
\begin{eqnarray}
{\cal E}_{s}^{\pm} &=& \sqrt{\frac{{\cal A}_{\pm}}{16\pi}} ~ \label{b18.1}
\end{eqnarray}
Next we compute rotational energy of ${\cal H}^{\pm}$ for this BH as
\begin{eqnarray}
{\cal E}_{r,\,J_{N}}^{\pm} &=& \sqrt{\frac{{\cal A}_{\pm}}{16\pi}+\frac{4\pi\,J_{N}^2}{{\cal A}_{\pm}}}
                       -\sqrt{\frac{{\cal A}_{\pm}}{16\pi}}.~ \label{b19}
\end{eqnarray}
Due to NUT parameter the electromagnetic energy  of ${\cal H}^{\pm}$ for RN-Taub-NUT BH becomes 
\begin{eqnarray}
{\cal E}_{em,\,J_{N}}^{\pm}
     &=& \sqrt{\frac{{\cal A}_{\pm}}{16\pi}+\frac{4\pi\,J_{N}^2}{{\cal A}_{\pm}}-N^2+\frac{4\pi\,N^4}{{\cal A}_{\pm}}} 
     -\sqrt{\frac{{\cal A}_{\pm}}{16\pi}+\frac{4\pi\,J_{N}^2}{{\cal A}_{\pm}}}.~ \label{b20}
\end{eqnarray}

Due to the charge parameter, the electromagnetic energy  of ${\cal H}^{\pm}$   for RN-Taub-NUT BH is
$$
{\cal E}_{em,\,J_{N},\,Q}^{\pm}= 
\sqrt{\frac{{\cal A}_{\pm}}{16\pi}+\frac{4\pi\,J_{N}^2}{{\cal A}_{\pm}}-N^2+\frac{4\pi\,N^4}{{\cal A}_{\pm}}
+\frac{\pi Q^4}{{\cal A}_{\pm}}-\frac{4\pi \,N^2\,Q^2}{{\cal A}_{\pm}}+\frac{Q^2}{2}}
$$
\begin{eqnarray}
-\sqrt{\frac{{\cal A}_{\pm}}{16\pi}+\frac{4\pi\,J_{N}^2}{{\cal A}_{\pm}}-N^2+\frac{4\pi\,N^4}{{\cal A}_{\pm}}} 
.~ \label{b21}
\end{eqnarray}
Next we compute the sum of these energies
$$
{\cal E}_{s}^{\pm}+{\cal E}_{r,\,J_{N}}^{\pm}+{\cal E}_{em,\,J_{N}}^{\pm}+{\cal E}_{em,\,J_{N},\,Q}^{\pm}=
$$
\begin{eqnarray}  
\sqrt{\frac{{\cal A}_{\pm}}{16\pi}+\frac{4\pi\,J_{N}^2}{{\cal A}_{\pm}}-N^2+\frac{4\pi\,N^4}{{\cal A}_{\pm}}
+\frac{\pi Q^4}{{\cal A}_{\pm}}-\frac{4\pi \,N^2\,Q^2}{{\cal A}_{\pm}}+\frac{Q^2}{2}}.
~ \label{b22}
\end{eqnarray}
Using Eq.~({\ref{b3}}), we can rewrite the above equation as
\begin{eqnarray}
{\cal E}_{s}^{\pm}+{\cal E}_{r,\,J_{N}}^{\pm}+\left({\cal E}_{em,\,J_{N}}^{\pm}+{\cal E}_{em,\,J_{N},\,Q}^{\pm}\right) 
&=&  M\,({\cal A}_{\pm}, J_{N}, N,Q) .~ \label{b23}
\end{eqnarray}

Remarkably, for RN-Taub-NUT BH, the Komar mass can also be expressed as the sum of these energies, namely 
the surface energy of ${\cal H}^{\pm}$, the rotational energy 
of ${\cal H}^{\pm}$, the electromagnetic energy  of ${\cal H}^{\pm}$ due to 
the NUT parameter and the electromagnetic energy  of ${\cal H}^{\pm}$ due 
to the charge parameter.  
Note that, in the limit $N=0$, we will get  the energy formula for RN BH i.e.
\begin{eqnarray}
{\cal E}_{s}^{\pm}+{\cal E}_{em,\,Q}^{\pm} &=& 
\sqrt{\frac{{\cal A}_{\pm}}{16\pi}+\frac{\pi Q^4}{{\cal A}_{\pm}}+\frac{Q^2}{2}}
=M\,({\cal A}_{\pm},Q).~ \label{b23.1}
\end{eqnarray}

\section{\label{3c} Energy Formula for Kerr-Taub-NUT BH}

The metric of Kerr-Taub-NUT BH~\cite{miller} in
Boyer-Lindquist like coordinates $(t, r, \theta, \phi)$ is given by 
\begin{eqnarray}
ds^2 &=& -\frac{\Delta}{\rho^2} \, \left[dt-P d\phi \right]^2+\frac{\sin^2\theta}{\rho^2}\,
\left[(r^2+a^2+N^2)\,d\phi-a\,dt\right]^2+\rho^2 \, \left[\frac{dr^2}{\Delta}+d\theta^2\right]
.\nonumber \\
~\label{c1}
\end{eqnarray}
where
\begin{eqnarray}
a &\equiv&\frac{J}{M},\, \rho^2 \equiv r^2+(N+a\,\cos\theta)^2 \\
\Delta &\equiv& r^2-2Mr+a^2-N^2\\
P & \equiv& a\,\sin^2\theta-2N\,\cos\theta  ~.\label{c2}
\end{eqnarray}
where the global conserved charges are the Komar mass $M$, angular momentum $J=a M$ and 
gravitomagnetic charge or dual mass or NUT parameter $N$. 

The radii of the horizon is evaluated by the function
$$\Delta|_{r=r_{\pm}}=0$$
which implies that
\begin{eqnarray}
r_{\pm} &\equiv&  M \pm\sqrt{ M^2-a^2+N^2}
\end{eqnarray}
Taking cognizance of  Eq.~({\ref{pn}}),  the area~\cite{wu} of ${\cal H}^\pm$ is thus 
\begin{eqnarray}
{\cal A}_{\pm}  &=& 8\pi\left[(M^2+N^2) \pm  \sqrt{M^4+J_{N}^2-J^2} \right] ~.\label{c3}
\end{eqnarray}

Analogously, the Komar mass  as a function of area, new conserved charges $J_{N}$, 
angular momentum~($J$) and NUT parameter for both the horizons~${\cal H}^{\pm}$:
\begin{eqnarray}
M\,({\cal A}_{\pm},\,J,\, J_{N},\, N) &=& 
\sqrt{\frac{{\cal A}_{\pm}}{16\pi}+\frac{4\pi\,J^2}{{\cal A}_{\pm}}+\frac{4\pi\,J_{N}^2}{{\cal A}_{\pm}}
-N^2+\frac{4\pi\,N^4}{{\cal A}_{\pm}}},
~\label{c4}
\end{eqnarray}
Similarly, the mass differential of  ${\cal H}^{\pm}$
\begin{eqnarray}
dM &=& {\cal T}_{\pm}\, d{\cal A}_{\pm}+\Omega_{\pm}\, dJ + \omega_{\pm}\, dJ_{N} +\psi_{\pm}\,dN ~. \label{c5}
\end{eqnarray}
where
\begin{eqnarray}
{\cal T}_{\pm} &=&  \frac{1}{{ M}} \left(\frac{1}{32 \pi}-\frac{2\pi J^2}{{\cal A}_{\pm}^2}
-\frac{2\pi J_{N}^2}{{\cal A}_{\pm}^2}-\frac{2\pi N^4}
{{\cal A}_{\pm}^2} \right),~\label{c6} \\
\omega_{\pm} &=& \frac{4\pi J_{N}}{M{\cal A}_{\pm}}, ~\label{c7} \\
\Omega_{\pm} &=& \frac{4\pi\,a}{{\cal A}_{\pm}}, ~\label{c8} \\
\psi_{\pm} &=& -\frac{N}{{M}} \left(1-\frac{8\pi N^2}{{\cal A}_{\pm}}\right)  
.~ \label{c9}
\end{eqnarray}
and
\begin{eqnarray}
{\cal T}_{\pm} &=& \mbox{Effective surface tension of ${\cal H}^\pm$}  \nonumber \\
\omega_{\pm} &=&  \mbox{Angular velocity of ${\cal H}^\pm$ due to NUT parameter} \nonumber \\
\Omega_{\pm} &=&  \mbox{Angular velocity of ${\cal H}^\pm$ due to spin parameter} \nonumber \\
\psi_{\pm} &=& \mbox{NUT potentials of ${\cal H}^\pm$}\nonumber
\end{eqnarray}
Similarly, the  effective surface tension is proportional to the surface gravity of ${\cal H}^\pm$
i.e.
\begin{eqnarray}
{\cal T}_{\pm} &=&  \frac{r_{\pm}-{M}}{8\pi\,\left(r_{\pm}^2+N^2+a^2 \right)}
= \frac{{\kappa}_{\pm}} {8\pi}, 
~\label{c10}
\end{eqnarray}
where $\kappa_{\pm}$ is defined as the surface gravity of ${\cal H}^{\pm}$ for Kerr-Taub-NUT BH. 
The BH temperature is 
\begin{equation}
T_{\pm}=\frac{\kappa_{\pm}}{ 2\pi}
=\frac{r_{\pm}-M}{2\pi\left(r_{\pm}^2+N^2+a^2\right)}.
~\label{c11}
\end{equation}

Similarly, the mass parameter of Kerr-Taub-NUT BH can be expressed in terms of these quantities 
as a simple bilinear form
\begin{eqnarray}
 M &=& 2{\cal T}_{\pm}\, {\cal A}_{\pm}+2J\, \Omega_{\pm}  + 2J_{N}\, \omega_{\pm} +\psi_{\pm}\, N
~. \label{c12}
\end{eqnarray}
This may be rewritten as 
\begin{eqnarray}
 M &=& \frac{{\kappa}_{\pm}} {4\pi}\, {\cal A}_{\pm}+2J\, \Omega_{\pm} 
 + 2J_{N}\, \omega_{\pm} +\psi_{\pm}\, N 
 ~. \label{c13}
\end{eqnarray}
Therefore  the \emph{Smarr-Gibbs-Duhem} relation of ${\cal H}^{\pm}$ for Kerr-Taub-NUT BH is 
\begin{eqnarray}
\frac{{M}}{2} &=& {T}_{\pm}{\cal S}_{\pm}++J\, \Omega_{\pm} + J_{N}\,\omega_{\pm}+\frac{\psi_{\pm}\,N}{2}
~. \label{c14}
\end{eqnarray}
As before, $dM$ is a perfect differential therefore one can choose freely 
any path of integration in $({\cal A}_{\pm},\,J, J_{N},\, N)$ space. So the 
surface energy  ${\cal E}_{s}^{\pm}$ of ${\cal H}^{\pm}$ for Kerr-Taub-NUT BH
can be defined by
\begin{eqnarray}
{\cal E}_{s}^{\pm} &=& \int_{0}^{{\cal A}_{\pm}} {\cal T}_{\pm} \,(\tilde{{\cal A}_{\pm}}
, 0 ,0,0)\, d\tilde{{\cal A}_{\pm}}; ~ \label{c15}
\end{eqnarray}
The rotational energy of ${\cal H}^{\pm}$ due to spin parameter can be defined by
\begin{eqnarray}
{\cal E}_{r,\,J}^{\pm} &=& \int_{0}^{J} \Omega_{\pm}\, ({\cal A}_{\pm}
, \tilde{J} ,0,0)\, d\tilde{J},\,\,  \mbox{${\cal A}_{\pm}$ fixed}; 
~ \label{c16}
\end{eqnarray}
The rotational energy  of ${\cal H}^{\pm}$ due to NUT parameter  can be defined by
\begin{eqnarray}
{\cal E}_{r,\,J,\,J_{N}}^{\pm} &=& \int_{0}^{J_{N}} \omega_{\pm}\, ({\cal A}_{\pm},J
,\tilde{J}_{N} ,0)\, d\tilde{J}_{N},\,\,  \mbox{${\cal A}_{\pm}$, $J$ fixed}; 
~ \label{c17}
\end{eqnarray}
and the electromagnetic energy  of ${\cal H}^{\pm}$ can be defined as 
\begin{eqnarray}
{\cal E}_{em,\,J,\,J_{N}}^{\pm} &=& \int_{0}^{N} \psi_{\pm} \,({\cal A}_{\pm}, J
, J_{N}, \tilde{N})\,d\tilde{N},\,\, \mbox{${\cal A}_{\pm}$, $J$, $J_{N}$  fixed};  
~ \label{c18}
\end{eqnarray}
These integrals should be directly computed using the variational definitions which is already 
defined in Eqs.~({\ref{c6}}), ({\ref{c7}}), ({\ref{c8}}), ({\ref{c9}}).
First, we shall compute the surface energy of ${\cal H}^{\pm}$ for Kerr-Taub-NUT BH
\begin{eqnarray}
{\cal E}_{s}^{\pm} &=& \sqrt{\frac{{\cal A}_{\pm}}{16\pi}} ~ \label{c19}
\end{eqnarray}
Next we find rotational energy of ${\cal H}^{\pm}$ due to spin parameter as
\begin{eqnarray}
{\cal E}_{r,\, J}^{\pm}  &=& \sqrt{\frac{{\cal A}_{\pm}}{16\pi}+\frac{4\pi\,J^2}{{\cal A}_{\pm}}}
                       -\sqrt{\frac{{\cal A}_{\pm}}{16\pi}}.~ \label{c20}
\end{eqnarray}
The rotational energy  of ${\cal H}^{\pm}$ due to NUT parameter is
\begin{eqnarray}
{\cal E}_{r,\,J,\,J_{N}}^{\pm} &=&  \sqrt{\frac{{\cal A}_{\pm}}{16\pi}+\frac{4\pi\,J^2}{{\cal A}_{\pm}}
+\frac{4\pi\,J_{N}^2}{{\cal A}_{\pm}}}-\sqrt{\frac{{\cal A}_{\pm}}{16\pi}+\frac{4\pi\,J^2}{{\cal A}_{\pm}}}
.~ \label{c20.1}
\end{eqnarray}
and finally we get the electromagnetic energy  of ${\cal H}^{\pm}$ for Kerr-Taub-NUT BH  as 
\begin{eqnarray}
{\cal E}_{em,\,J\,J_{N}}^{\pm} &=& 
\sqrt{\frac{{\cal A}_{\pm}}{16\pi}+\frac{4\pi\,J^2}{{\cal A}_{\pm}}+\frac{4\pi\,J_{N}^2}{{\cal A}_{\pm}}-N^2
+\frac{4\pi\,N^4}{{\cal A}_{\pm}}} -\sqrt{\frac{{\cal A}_{\pm}}{16\pi}+\frac{4\pi\,J^2}{{\cal A}_{\pm}}
+\frac{4\pi\,J_{N}^2}{{\cal A}_{\pm}}}.\nonumber \\
~ \label{c21}
\end{eqnarray}
Now we compute the sum of these energies 
\begin{eqnarray}
{\cal E}_{s}^{\pm}+{\cal E}_{r,\,J}^{\pm}+{\cal E}_{r,\,J,\,J_{N} }^{\pm}+{\cal E}_{em,\,J,\,J_{N}}^{\pm} &=&  
\sqrt{\frac{{\cal A}_{\pm}}{16\pi}+\frac{4\pi\,J^2}{{\cal A}_{\pm}}
+\frac{4\pi\,J_{N}^2}{{\cal A}_{\pm}}-N^2+\frac{4\pi\,N^4}{{\cal A}_{\pm}}}.\nonumber \\
~ \label{c22}
\end{eqnarray}
Using Eq.~({\ref{c4}}), we can rewrite the above equation for Kerr-Taub-NUT BH as
\begin{eqnarray}
{\cal E}_{s}^{\pm}+\left({\cal E}_{r,\,J}^{\pm}+{\cal E}_{r,\,J,\,J_{N}}^{\pm}\right)
+{\cal E}_{em,\,J,\,J_{N}}^{\pm} &=&  M\,({\cal A}_{\pm},\,J,\,J_{N},\,N) .~ \label{c23}
\end{eqnarray}

Remarkably for Kerr-Taub-NUT BH, the Komar mass can be expressed as sum of four
energies namely the surface energy of ${\cal H}^{\pm}$, the rotational energy of 
${\cal H}^{\pm}$ due to spin parameter, the rotational energy  of ${\cal H}^{\pm}$ due 
to NUT parameter and the electromagnetic energy  of ${\cal H}^{\pm}$. Note that 
in the limit $N=0$, one obtains the energy formula for Kerr BH i.e. 
\begin{eqnarray}
{\cal E}_{s}^{\pm}+{\cal E}_{r,\,J}^{\pm} &=&  
\sqrt{\frac{{\cal A}_{\pm}}{16\pi}+\frac{4\pi\,J^2}{{\cal A}_{\pm}}}=
M\,({\cal A}_{\pm},\,J) .~ \label{c23.1}
\end{eqnarray}

\section{\label{3d} Energy Formula for Kerr-Newman-Taub-NUT BH}
Finally, we consider most general class of BH without cosmological constant. 
The metric form similar to Eq.~(\ref{c1}) and having the horizon 
function~\cite{miller} as
\begin{eqnarray}
\Delta &\equiv&  r^2-2 M r+a^2+Q^2-N^2  ~.\label{delknn}
\end{eqnarray}
The horizons are located at 
\begin{eqnarray}
r_{\pm} &\equiv &  M \pm \sqrt{M^2-a^2-Q^2+N^2}
\end{eqnarray}
Analogously, the BH horizon area of ${\cal H}^\pm$ is 
\begin{eqnarray}
{\cal A}_{\pm} &=& 4\pi\left[2(M^2+N^2)-Q^2 \pm 2 \sqrt{M^4+J_{N}^2-J^2-M^2Q^2} \right]
~.\label{d1}
\end{eqnarray}
Analogously the mass parameter of Kerr-Newman-Taub-NUT BH for both the 
horizons~${\cal H}^{\pm}$ are
$$
M\,({\cal A}_{\pm},\,J,\, J_{N},\, N,\, Q) =
$$
\begin{eqnarray}
\sqrt{\frac{{\cal A}_{\pm}}{16\pi}+\frac{4\pi\,J^2}{{\cal A}_{\pm}}
+\frac{4\pi\,J_{N}^2}{{\cal A}_{\pm}}-N^2+\frac{4\pi\,N^4}{{\cal A}_{\pm}}
+\frac{\pi Q^4}{{\cal A}_{\pm}}-\frac{4\pi \,N^2\,Q^2}{{\cal A}_{\pm}}+\frac{Q^2}{2}},
~\label{d3}
\end{eqnarray}
Hence  the mass differential of ${\cal H}^{\pm}$ becomes
\begin{eqnarray}
dM &=& {\cal T}_{\pm}\, d{\cal A}_{\pm}+\Omega_{\pm}\,dJ + \omega_{\pm}\, dJ_{N} +\psi_{\pm}\,dN+\Phi_{\pm}\,dQ
~. \label{d4}
\end{eqnarray}
where
\begin{eqnarray}
{\cal T}_{\pm} &=&  \frac{1}{{M}} \left(\frac{1}{32 \pi}-\frac{2\pi J_{N}^2}{{\cal A}_{\pm}^2}-\frac{2\pi N^4}
{{\cal A}_{\pm}^2}-\frac{2\pi J^2}{{\cal A}_{\pm}^2}-\frac{\pi\,Q^4}{2{\cal A}_{\pm}^2}+\frac{2\pi \,N^2\,Q^2}
{{\cal A}_{\pm}^2} \right)~.\label{d5} \\
\omega_{\pm} &=& \frac{4\pi J_{N}}{M{\cal A}_{\pm}}, ~.\label{d6} \\
\Omega_{\pm} &=& \frac{4\pi\,a}{{\cal A}_{\pm}}, ~\label{d8} \\
\psi_{\pm} &=& -\frac{N}{{M}} \left(1-\frac{8\pi N^2}{{\cal A}_{\pm}}+\frac{4\pi\,Q^2}{{\cal A}_{\pm}}\right)  ~. \label{d9}\\
\Phi_{\pm} &=& \frac{1}{M} \left(\frac{2\pi\,Q^3}{{\cal A}_{\pm}}-\frac{4\pi\,Q\,N^2}{{\cal A}_{\pm}}+\frac{Q}{2} \right)
 ~. \label{d10}
\end{eqnarray}
and
\begin{eqnarray}
{\cal T}_{\pm} &=& \mbox{Effective surface tension of ${\cal H}^{\pm}$ for KN-Taub-NUT BH} \nonumber \\
\omega_{\pm} &=&  \mbox{Angular velocity of ${\cal H}^\pm$ due to NUT parameter} \nonumber \\
\Omega_{\pm} &=&  \mbox{Angular velocity of ${\cal H}^\pm$ due to spin parameter} \nonumber \\
\psi_{\pm} &=& \mbox{Electromagnetic potentials of ${\cal H}^\pm$ due to NUT parameter}\nonumber\\
\Phi_{\pm} &=& \mbox{Electromagnetic potentials of ${\cal H}^\pm$ due to spin parameter}\nonumber
\end{eqnarray}

Similarly, we prove that this effective surface tension is proportional to the surface gravity 
of the horizons, i.e.
\begin{eqnarray}
{\cal T}_{\pm} &=& \frac{r_{\pm}-{M}}{8\pi\,\left(2 M r_{\pm}+2N^2-Q^2 \right)}
                =\frac{r_{\pm}-{M}}{8\pi\,\left(r_{\pm}^2+N^2+a^2 \right)}
            =\frac{{\kappa}_{\pm}} {8\pi},~\label{d11}
\end{eqnarray}
where $\kappa_{\pm}$ is defined {as} the surface gravity of ${\cal H}^{\pm}$ for KN-Taub-NUT BH. 

Analogously, the bilinear form of the mass parameter is
\begin{eqnarray}
 M &=& 2{\cal T}_{\pm}\, {\cal A}_{\pm}+2J\, \Omega_{\pm}+
 2J_{N}\, \omega_{\pm} +\psi_{\pm}\, N+\Phi_{\pm}\, Q
~. \label{d12}
\end{eqnarray}
Like Kerr-Taub-NUT BH, this striking formula may be obtained by 
applying Euler's theorem on homogeneous functions to $M$, which 
is homogeneous of degree 
$\frac{1}{2}$ in $({\cal A}_{\pm},\, J,\,J_{N},\, N^2, Q^2)$.
Thus it can be rewritten as 
\begin{eqnarray}
 M &=& \frac{{\kappa}_{\pm}} {4\pi}\, {\cal A}_{\pm}+2J\, \Omega_{\pm} 
 + 2J_{N}\, \omega_{\pm} +\psi_{\pm}\, N+\Phi_{\pm}\,Q
 ~. \label{d13}
\end{eqnarray}
So,  the \emph{Smarr-Gibbs-Duhem} relation of ${\cal H}^{\pm}$ for 
Kerr-Newman-Taub-NUT BH is 
\begin{eqnarray}
\frac{{M}}{2} &=& {T}_{\pm}{\cal S}_{\pm}+J\, \Omega_{\pm}+ J_{N}\,\omega_{\pm}
+\frac{\psi_{\pm}\,N}{2}+\frac{\Phi_{\pm}\,Q}{2}
~. \label{d14}
\end{eqnarray}
Interestingly, ${\cal T}_{\pm}$,  $\omega_{\pm}$, $\psi_{\pm}$ and $\Phi_{\pm}$ should be  
defined and are constant on  ${\cal H}^{\pm}$.

Proceeding similarly, the surface energy 
${\cal E}_{s}^{\pm}$ of ${\cal H}^{\pm}$ for this BH can be defined by
\begin{eqnarray}
{\cal E}_{s}^{\pm} &=& \int_{0}^{{\cal A}_{\pm}} {\cal T}_{\pm} \,(\tilde{{\cal A}_{\pm}}
, 0 ,0,0,0)\, d\tilde{{\cal A}_{\pm}}; ~ \label{d15}
\end{eqnarray}
The rotational energy of ${\cal H}^{\pm}$ due to spin parameter can be defined by
\begin{eqnarray}
{\cal E}_{r,\,J}^{\pm} &=& \int_{0}^{J} \Omega_{\pm}\, ({\cal A}_{\pm}
, \tilde{J}, 0, 0, 0)\, d\tilde{J},\,\,  \mbox{${\cal A}_{\pm}$ fixed}; 
~ \label{d16}
\end{eqnarray}
The rotational energy  of ${\cal H}^{\pm}$ due to NUT parameter  can be defined by
\begin{eqnarray}
{\cal E}_{r,\,J,\,J_{N}}^{\pm} &=& \int_{0}^{J_{N}} \omega_{\pm}\, ({\cal A}_{\pm},J
,\tilde{J}_{N}, 0, 0)\, d\tilde{J}_{N},\,\,  \mbox{${\cal A}_{\pm}$, $J$ fixed}; 
~ \label{d17}
\end{eqnarray}
The electromagnetic energy  of ${\cal H}^{\pm}$ due to NUT parameter~($N$) can be defined as 
\begin{eqnarray}
{\cal E}_{em,\,J,\,J_{N}}^{\pm} &=& \int_{0}^{N} \psi_{\pm} \,({\cal A}_{\pm}, J
, J_{N}, \tilde{N},0)\,d\tilde{N},\,\, \mbox{${\cal A}_{\pm}$, $J$, $J_{N}$  fixed};  ~ \label{d17.1}
\end{eqnarray}
and the electromagnetic energy  of ${\cal H}^{\pm}$ due to charge parameter~($Q$) 
can be defined as 
\begin{eqnarray}
{\cal E}_{em,\,J,\,J_{N},\,Q}^{\pm} &=& \int_{0}^{Q} \Phi_{\pm} \,({\cal A}_{\pm}, J
, J_{N}, N, \tilde{Q})\, d\tilde{Q},\,\, \mbox{${\cal A}_{\pm}$, $J$, $J_{N}$, $N$ fixed};  ~ \label{d18}
\end{eqnarray}
These integrals may be directly computed using the variational definitions which are
defined in Eqs.~({\ref{d5}}), ({\ref{d6}}), ({\ref{d8}}), ({\ref{d9}}), ({\ref{d10}}).

Proceeding similarly, the surface energy of ${\cal H}^{\pm}$ for Kerr-Newman-Taub-NUT BH is
\begin{eqnarray}
{\cal E}_{s}^{\pm} &=& \sqrt{\frac{{\cal A}_{\pm}}{16\pi}} ~ \label{d20}
\end{eqnarray}
The rotational energy  of ${\cal H}^{\pm}$ due to  spin parameter
\begin{eqnarray}
{\cal E}_{r,\,J}^{\pm}  &=& \sqrt{\frac{{\cal A}_{\pm}}{16\pi}+\frac{4\pi\,J^2}{{\cal A}_{\pm}}}
                       -\sqrt{\frac{{\cal A}_{\pm}}{16\pi}}.~ \label{d22}
\end{eqnarray}
The rotational energy  of ${\cal H}^{\pm}$ due to  NUT parameter is
\begin{eqnarray}
{\cal E}_{r,\,J,\, J_{N}}^{\pm} &=&  \sqrt{\frac{{\cal A}_{\pm}}{16\pi}+\frac{4\pi\,J^2}{{\cal A}_{\pm}}
+\frac{4\pi\,J_{N}^2}{{\cal A}_{\pm}}}-\sqrt{\frac{{\cal A}_{\pm}}{16\pi}+\frac{4\pi\,J^2}{{\cal A}_{\pm}}}
.~ \label{d23}
\end{eqnarray}

The electromagnetic energy  of ${\cal H}^{\pm}$ due to NUT parameter is  
\begin{eqnarray}
{\cal E}_{em,\,J,\,J_{N}}^{\pm}
     &=& \sqrt{\frac{{\cal A}_{\pm}}{16\pi}+\frac{4\pi\,J^2}{{\cal A}_{\pm}}
     +\frac{4\pi\,J_{N}^2}{{\cal A}_{\pm}}-N^2+\frac{4\pi\,N^4}{{\cal A}_{\pm}}} 
     -\sqrt{\frac{{\cal A}_{\pm}}{16\pi}+\frac{4\pi\,J^2}{{\cal A}_{\pm}}
     +\frac{4\pi\,J_{N}^2}{{\cal A}_{\pm}}}.\nonumber \\
     ~ \label{d24}
\end{eqnarray}

The electromagnetic energy  of ${\cal H}^{\pm}$ due to the charge parameter is
$$
{\cal E}_{em,\,J,\,J_{N},\,Q}^{\pm}= 
\sqrt{\frac{{\cal A}_{\pm}}{16\pi}+\frac{4\pi\,J_{N}^2}{{\cal A}_{\pm}}-N^2+\frac{4\pi\,N^4}{{\cal A}_{\pm}}+
\frac{4\pi\,J^2}{{\cal A}_{\pm}}
+\frac{\pi Q^4}{{\cal A}_{\pm}}-\frac{4\pi \,N^2\,Q^2}{{\cal A}_{\pm}}+\frac{Q^2}{2}}
$$
\begin{eqnarray}
-\sqrt{\frac{{\cal A}_{\pm}}{16\pi}+\frac{4\pi\,J^2}{{\cal A}_{\pm}}
+\frac{4\pi\,J_{N}^2}{{\cal A}_{\pm}}-N^2+\frac{4\pi\,N^4}{{\cal A}_{\pm}}} 
.~ \label{d25}
\end{eqnarray}
Next we compute the sum of these energies 
$$
{\cal E}_{s}^{\pm}+{\cal E}_{r,\,J}^{\pm}+{\cal E}_{r,\,J,\, J_{N}}^{\pm}
+{\cal E}_{em,\,J,\,J_{N}}^{\pm}+{\cal E}_{em,\,J,\,J_{N},\,Q}^{\pm} 
= 
$$
\begin{eqnarray}
\sqrt{\frac{{\cal A}_{\pm}}{16\pi}+\frac{4\pi\,J^2}{{\cal A}_{\pm}}
+\frac{4\pi\,J_{N}^2}{{\cal A}_{\pm}}-N^2+\frac{4\pi\,N^4}{{\cal A}_{\pm}}
+\frac{\pi Q^4}{{\cal A}_{\pm}}-\frac{4\pi \,N^2\,Q^2}{{\cal A}_{\pm}}+\frac{Q^2}{2}}.
~ \label{d25.1}
\end{eqnarray}

Using Eq.~({\ref{d3}}), we can rewrite the above equation as
\begin{eqnarray}
{\cal E}_{s}^{\pm}+\left({\cal E}_{r,\, J}^{\pm}+{\cal E}_{r,\,J,\, J_{N}}^{\pm}\right)
+\left({\cal E}_{em,\,J,\,J_{N}}^{\pm}+{\cal E}_{em,\,J,\,J_{N},\,Q}^{\pm}\right) 
&=&  M\,({\cal A}_{\pm}, J, J_{N}, N, Q).\nonumber\\
~ \label{d26}
\end{eqnarray}

Remarkably, for KN-Taub-NUT BH the Komar mass can also be expressed as 
sum of these energies namely the surface energy of ${\cal H}^{\pm}$, 
the rotational energy of ${\cal H}^{\pm}$, the electromagnetic energy of 
${\cal H}^{\pm}$ due to NUT parameter and the electromagnetic energy of 
${\cal H}^{\pm}$ due to charge parameter. It should be noted  
that in the limit $N=0$, one obtains the energy formula for Kerr-Newman BH~\cite{smarr73a}
i.e. 
\begin{eqnarray}
{\cal E}_{s}^{\pm}+{\cal E}_{r,\, J}^{\pm}+{\cal E}_{em,\,J,\,Q}^{\pm} &=&  
\sqrt{\frac{{\cal A}_{\pm}}{16\pi}+\frac{4\pi\,J^2}{{\cal A}_{\pm}}
+\frac{\pi Q^4}{{\cal A}_{\pm}}+\frac{Q^2}{2}}
=M\,({\cal A}_{\pm}, J, Q) .~ \label{d26.1}
\end{eqnarray}

\section{\label{con} Conclusions}
It has been suggested that a generic four {dimensional} Taub-NUT BH~\cite{wu} should be 
completely specified in terms of three or four different types of thermodynamic hairs. 
They are  defined as the Komar mass~($M=m$), the angular momentum~($J_{n}=mn$), 
the gravitomagnetic charge ($N=n$),  the dual~(magnetic) mass $(\tilde{M}=n)$. 
Under this formalism, we derived the surface energy of ${\cal H}^{\pm}$, the 
rotational energy of ${\cal H}^{\pm}$ and the electromagnetic energy of ${\cal H}^{\pm}$ for NUT class of BHs.
Interestingly, we showed that like Kerr-Newman BH~\cite{smarr73a,pp14} the mass parameter could be 
expressed as the sum of these three energies i. e. 
$M={\cal E}_{s}^{\pm}+{\cal E}_{r}^{\pm}+{\cal E}_{em}^{\pm}$. 
We have explicitly examined for Taub-NUT BH, RN-Taub-NUT BH, 
Kerr-Taub-NUT BH and KN-Taub-NUT BH. In each case of the NUT class of BHs, we showed that
the sum of these three energies is equal to the Komar mass. It has been achieved only due to the 
introduction of new conserved charges~$J_{N}=M\,N$, which is closely analogous to Kerr-like angular 
momentum parameter~($J=a\,M$).

\bibliography{apssamp}

\end{document}